\documentclass{wsp-mb}

\begin{document}

\title{A Glimpse at Mathematical Diffraction Theory}

\author{Michael Baake}

\address{Institut f\"ur Mathematik, Universit\"at Greifswald \\
Jahnstr.~15 a, 17487 Greifswald, Germany \\
Email: {\sf mbaake@uni-greifswald.de}}

\maketitle

\abstracts{
Mathematical diffraction theory is concerned with the analysis
of the diffraction measure of a translation bounded complex 
measure $\omega$. It emerges as the Fourier transform of the 
autocorrelation measure of $\omega$.
The mathematically rigorous approach has produced a number
of interesting results in the context of perfect and random
systems, some of which are summarized here. 
}

\thispagestyle{empty}

\section{Introduction}
Diffraction experiments are widely used to gain information on the
atomic structure of solids. Neutron and $X$-ray diffraction share
the property that fundamental aspects of the diffraction image can
be understood very well in single scattering theory together with Born's 
approximation, while electron diffraction needs more sophistication,
see Ref.~\raisebox{-1.2ex}{\Large \cite{Cowley}} for details. 
In the former case, on which we will
focus here, the diffraction image is then described as the Fourier
transform of the autocorrelation of the original distribution of
scatterers. In this situation, important information is lost,
namely the relative phases. Consequently, the inverse problem of
structure determination has, in general, no unique solution, unless
extra information is available from other sources of knowledge.

All this can be phrased in mathematically rigorous terms, but hardly
anybody seemed really interested in it -- until quasicrystals were
discovered in the early 1980's. They are solids which, like crystals,
show sharp Bragg diffraction images in $X$-ray diffraction, yet with
maliciously perfect point symmetries which are crystallographically
forbidden -- at least in dimensions $\le 3$. As turned out quickly,
they can be formally described as projections of portions (strips)
of lattices in higher dimensions, and this can account both for their
symmetries and for their perfect Bragg diffraction spectra. However,
they also leave a certain degree of uneasiness towards the obviously
not sufficiently well understood possibilities in diffraction theory.

This was the starting point of the rigorous development of mathematical
diffraction theory in a paper by Hof~\cite{Hof1} which was later also
extended to the treatment of high temperature.\cite{Hof2}
The setting is as follows. The
structure under investigation is described as a 
translation bounded complex measure $\omega$ on $n$-dimensional 
Euclidean space, whose autocorrelation, $\gamma^{}_{\omega}$
is obtained as a volume-weighted convolution,
\begin{equation} \label{auto1}
   \gamma^{}_{\omega} \; = \; \lim_{r\to\infty}\,
   \frac{\omega|_r * \tilde{\omega}|_r}{{\rm vol}(B_r)} 
\end{equation}
where $B_r$ is the (open) ball of radius $r$ around $0$,
$\omega|_r$ denotes the restriction of $\omega$ to $B_r$, and
$\,\tilde{}\,$ stands for taking the complex conjugate of the
origin-inverted measure. It is not even clear in general whether
this limit exists, but if not, the right hand side certainly has points
of accumulation which can then be obtained as limits along a discrete 
subsequence of averaging balls. We will thus assume, from now on, that 
the limit in (\ref{auto1}) exists.

If so, the autocorrelation is a positive definite measure, and as such
automatically possesses a Fourier transform, denoted by 
$\hat{\gamma}^{}_{\omega}$, which is a positive measure by Bochner's 
theorem, compare Ref.~\raisebox{-1.2ex}{\Large \cite{BF}} for
background material. Furthermore, $\hat{\gamma}^{}_{\omega}$ has a unique
decomposition into three parts,
\begin{equation}  \label{decomp}
  \hat{\gamma}^{}_{\omega} \; = \; \big(\hat{\gamma}^{}_{\omega}\big)_{pp} +
  \big(\hat{\gamma}^{}_{\omega}\big)_{sc} + 
  \big(\hat{\gamma}^{}_{\omega}\big)_{ac} \, ,
\end{equation}
where $pp$, $sc$ and $ac$ as usual stand for pure point, singular continuous
and absolutely continuous, with the ordinary Lebesgue measure as reference.
The pure point (absolutely continuous) part corresponds to the Bragg spectrum
(to diffuse scattering), while the term `singular continuous' does not appear
in the standard crystallographic literature.

In what follows, we will summarize a few aspects of this setting and give
some pointers for further reading.

\section{Pure point diffraction}

One important class of structures are those with pure Bragg spectrum, i.e.\
those where $\hat{\gamma}^{}_{\omega}$ is a pure point measure. This is
certainly the case if the measure $\omega$ is crystallographic, i.e.\ periodic
with a full lattice $\Gamma$ of periods. So, $\omega = h * \delta^{}_{\Gamma}$
where $h$ is the density (or measure) in a fundamental domain and
$\delta^{}_{\Gamma}$ is the uniform lattice Dirac comb. With this, the
autocorrelation reads 
$\gamma^{}_{\omega} = (h*\tilde{h})\, {\rm dens}(\Gamma)\, \delta^{}_{\Gamma}$,
where ${\rm dens}(\Gamma)$ is the density of the lattice, and the 
diffraction measure is 
\begin{equation}
  \hat{\gamma}^{}_{\omega} \; = \; |\hat{h}|^2 \, 
   \big( {\rm dens}(\Gamma)\big)^2 \, \delta^{}_{\Gamma^*}
\end{equation}
with $\Gamma^* = \{u\mid uv \in {\sf Z\hspace*{-.9ex}Z} \mbox{ for all } v\in\Gamma\}$
the dual lattice. This rests upon Poisson's summation formula,
$\hat{\delta}^{}_{\Gamma} = {\rm dens}(\Gamma) \delta^{}_{\Gamma^*}$,
and reconfirms that the diffraction spectrum of a perfect crystal is
pure point.

Other by now well-known examples of pure point diffractive systems include
perfect quasicrystals as described by the projection method. More precisely,
given a cut and project scheme,\cite{Moody} a generic, regular model set 
based on it will produce a pure point diffraction spectrum. Here, a subset
of a lattice is projected which lies in an incommensurately oriented strip
with constant cross section. The latter is called window and lives in
the so-called internal space, while the complement, which the pattern
is projected to, is termed direct or physical space. The diffraction 
result was first proved by Hof~\cite{Hof1} for Euclidean internal 
spaces with polytopal windows. It was later extended by various people
to include more general internal spaces, and finally proved in full generality 
by Schlottmann.\cite{Martin} This fuelled the speculation that basically 
only model sets have this property, but the set of visible lattice 
points~\cite{BMP} proved otherwise. In fact, model sets only form a 
relatively sparse set of examples with pure point diffraction.

After some effort into solving this puzzle, the answer uses a different
box of tools and relates diffraction properties to almost periodicity of
the autocorrelation measure. Partly based on results of Gil de Lamadrid
and Argabright, it is shown in Ref.~\raisebox{-1.2ex}{\Large \cite{BM}} 
that a translation bounded measure
$\omega$ has pure point diffraction if and only if its autocorrelation
$\gamma^{}_{\omega}$ is strongly almost periodic as a measure. This criterion
can actually be checked in many cases, and it also provides an independent
proof of Schlottmann's theorem as a by-product. The result also shows that
the diffraction problem really consists of two steps: one from $\omega$ to
its autocorrelation, $\gamma^{}_{\omega}$, and one from there to the Fourier 
transform, $\hat{\gamma}^{}_{\omega}$. The second step is ultimately easy
because it is at least one-to-one, even though many open questions are still
to be settled. So, let us take a look at the first step, which lies at the
heart of the phase reconstruction problem.

\section{The homometry problem}

Two different measures can share the same autocorrelation, as is obvious
from the limit in (\ref{auto1}), in which case they are called
{\em homometric}. In particular, modifying a given measure
by a {\em finite\/} measure has no effect on the limit, i.e.\ on
$\gamma^{}_{\omega}$. But much worse operations are possible, which
makes the equivalence relation defined by homometry a rather nasty one.
To show how bad the situation is, we recall a result from 
Ref.~\raisebox{-1.2ex}{\Large \cite{HB}}.
There, the binary Rudin-Shapiro sequence, which is based on a substitution
rule and thus has entropy $0$, is realized as a Dirac comb along the
integers. It has the same autocorrelation as the generic Bernoulli
Dirac comb with occupation probability $1/2$ per site, which is truly
random and has maximal entropy, $\log(2)$. The homometry class also 
contains an example for each intermediate value of the entropy. This
shows how bad the inverse problem really is, unless further information
is available about the structure, or a general principle is invoked to
select a `reasonable' representative from the homometry class, e.g.\
by maximizing the entropy.

Another example is discussed in Ref.~\raisebox{-1.2ex}{\Large \cite{MB2001}}. 
A subset $S$ of a lattice is
homometric with its complement set, if both have the same density --
no matter whether the subset itself is periodic, aperiodic or random.
So, one can in particular construct homometric pairs for all possible
types of spectra, pure and mixed. Note that the pure point part which
is due to the underlying lattice can be made extinct by simply
subtracting a multiple of the uniform lattice Dirac comb. This way,
one can even realize purely singular continuous spectra, e.g.\ based
upon substitution rules of Thue-Morse type.

\section{Diffraction of lattice subsets}

A subset of a lattice  $\Gamma$ inherits a good deal of the lattice structure, 
no matter how unusual the selection of lattice points may be. In particular,
if we once again make the (rather weak) assumption that a given subset $S$
has a unique autocorrelation, the diffraction measure 
$\hat{\gamma}^{}_{\omega}$ of the Dirac comb $\delta^{}_S$ is periodic,
with the dual lattice $\Gamma^*$ as lattice of periods.\cite{MB2001}
What is more, the autocorrelation admits a representation in the form
$\gamma^{}_{\omega} = \Phi \cdot \delta^{}_{\Gamma}$ with a Lipschitz
continuous function $\Phi$ which interpolates the autocorrelation
coefficients between the discrete lattice points. It can actually
be chosen to have an extension to an entire function, which is a
bit surprising.

In connection with this, the diffraction measure takes the form
\begin{equation}  \label{period}
  \hat{\gamma}^{}_{\omega} \; = \; \varrho * \delta^{}_{\Gamma^*}
\end{equation}
where $\varrho$ is a finite, positive measure concentrated on a
true fundamental domain of the dual lattice, see 
Ref.~\raisebox{-1.2ex}{\Large \cite{MB2001}} for details. This
means that the analysis of the spectral type can be done with
$\varrho$, which makes it considerably easier. 

This result, like many mentioned so far, can be generalized 
substantially, e.g.\ to weighted lattice subsets or to 
the setting of lattices in locally compact 
Abelian groups that are $\sigma$-compact. On the other hand, it
has a wide range of applications. First of all, the above mentioned
set of visible lattice points falls into the class, so must have
a diffraction of the convolution form (\ref{period}), as was
previously derived explicitly.\cite{BMP} Another class of interesting
point sets are those obtained from lattice substitution systems
which have recently received a lot of attention.\cite{LM}
{}Finally, all lattice gases, with or without interaction, are covered.
This gives a unified explanation for the periodicity observed in
the diffraction of these stochastic systems.

\section{Diffraction from stochastic point sets}

Let us finally take a closer look at systems with some degree of
randomness, either due to a lattice gas structure or due to an
underlying random tiling approach. Bernoulli subsets of lattices
and model sets are well understood, even in terms of elementary
methods from stochastics, see 
Ref.~\raisebox{-1.2ex}{\Large \cite{BM-random}}
and references therein. Recently, this approach has been generalized
considerably, with more sophisticated machinery, to cover stochastic
selections from rather general Delone sets.\cite{K}
The outcome proves the folklore claim that uncorrelated random
removal of scatterers has two effects, namely reducing the overall
intensity of the diffraction of the fully occupied set, without
changing the relative intensities, and adding a white noise type
constant diffuse background.

More interesting would be a detailed understanding of the spectra
of stochastic systems with interactions. Though some qualitative
features are known, quantitative results are rare, i.e.\ basically
restricted to systems in one dimension or to systems that can be mapped
to exactly solvable models of statistical mechanics, see 
Ref.~\raisebox{-1.2ex}{\Large \cite{BH}} and the literature given 
there. A particularly interesting question concerns the precise
spectral nature of random tilings of the plane, both with
crystallographic and non-crystallographic symmetries (which is
essentially meant as symmetry on average here). While typical dimer
based lattice random tilings have mixed spectra with pure point
and absolutely continuous components,\cite{BH} some quasicrystal related
random tilings seem to have singular continuous parts. There are
very convincing scaling arguments and numerical simulations in
favour of this observation, but no proof is available at present.
This is an open challenge, together with the investigation how
robust and hence relevant such a singular continuous diffraction
spectrum would be for real world crystallography.

\section*{Acknowledgments}

This summary grew out of discussions and cooperations with a number of
colleagues, in particular Moritz H\"offe, Robert V.\ Moody and Martin
Schlottmann, whose input is gratefully acknowledged. I thank Uwe Grimm
for reading the manuscript and suggesting a number of improvements.

\end{document}